\documentclass[aps,prl,twocolumn,showpacs]{revtex4}

\usepackage{graphicx}
\usepackage{bm}
\usepackage{amsmath}

\begin{document}


\title{Microscopic Calculation of Spin Torques in 
Disordered Ferromagnets }
\author{Hiroshi Kohno}
\email{kohno@mp.es.osaka-u.ac.jp}
\affiliation{
Graduate School of Engineering Science, Osaka University,
Toyonaka, Osaka 560-8531, Japan}
\author{Gen Tatara}
\affiliation{
Graduate School of Science, Tokyo Metropolitan University,
Hachioji, Tokyo 192-0397, Japan \\
PRESTO, JST, Kawaguchi, Saitama 332-0012, Japan}
\author{Junya Shibata}
\affiliation{RIKEN-FRS, 2-1 Hirosawa, 
Wako, Saitama 351-0198, Japan}

\date{\today}

\begin{abstract}
 Effects of conduction electrons on magnetization dynamics, 
represented by spin torques, are calculated microscopically 
in the first order in spatial gradient and time derivative of 
magnetization. 
 Special attention is paid to the so-called $\beta$-term and the 
Gilbert damping, $\alpha$, in the presence of electrons' spin-relaxation 
processes, 
which are modeled by quenched magnetic (and spin-orbit) impurities. 
 The obtained results such as $\alpha \ne \beta$ hold for 
localized as well as itinerant ferromagnetism. 
\end{abstract}
\maketitle

Recent activity in spintronics to manipulate nanoscale magnetization, 
such as magnetization reversal, by electric current started with the 
concept of spin-transfer effect introduced 
by Slonczewski \cite{Slonczewski} and Berger \cite{Berger96}. 
 When a current flows in a background magnetization ${\bm n}$ varying in 
space, a slight directional mismatch (with ${\bm n}$) arises 
in the electron spin polarization 
such that spin angular momentum of a spin-polarized current (spin current) 
is transferred to the magnetization. 
 This effect is expressed as a torque on ${\bm n}$, 
called spin-transfer torque; 
for slowly-varying ${\bm n}$, it takes the form, 
$-({\bm v}_{\rm s} \!\cdot\! {\bm \nabla}) {\bm n}$, 
where ${\bm v}_{\rm s}$ is a velocity characterizing the spin-transfer rate 
\cite{BJZ98,Ansermet04,Macdonald04,Li04,STK05}.

 Existence of another type of current-induced spin torque, of the form, 
$- \beta {\bm n} \times ({\bm v}_{\rm s} \!\cdot\! {\bm \nabla}) {\bm n}$, 
was noted recently \cite{Zhang05,Thiaville05,Barnes05}.
 This torque is perpendicular to the spin-transfer torque, 
and is called nonadiabatic torque or the $\beta$-term. 
 Zhang and Li showed on phenomenological grounds 
that it arises as a result of spin relaxation of conduction electrons 
\cite{Zhang05}. 
 Barnes and Maekawa suggested its close connection to the Gilbert damping, 
$\alpha$, and proposed a relation $\alpha = \beta$ \cite{Barnes05}. 
 While the importance of the $\beta$-term in magnetization dynamics 
is now well-recognized \cite{Zhang05,Thiaville05,Barnes05,TTKSNF06}, 
its microscopic derivation is a current issue.

 Similar torques are known also in multilayer 
\cite{Heide01,multilayer_review}, 
domain wall \cite{Berger84,WV,TK04}, 
and other \cite{Edwards05,Onoda05} systems. 
 For example, for a rigid domain wall, a torque which acts as a force and 
may thus be called momentum-transfer torque, was noted \cite{Berger84,TK04}. 
 This is a spatially oscillating torque 
due to electron reflection, found by Waintal and Viret \cite{WV}, 
which has the same algebraic form as the $\beta$-term but is 
spatially nonlocal \cite{KTS}. 
 This torque can be important for magnetic configurations varying 
rapidly in space whereas the $\beta$-term is relevant to slowly-varying 
configurations. 
 We focus on the $\beta$-term in this Letter.

 Very recently, a microscopic study of this subjest was undertaken 
by Tserkovnyak, Brataas and Bauer (TBB) 
based on the Boltzmann equation \cite{TBB}. 
 They showed that $\alpha = \beta$ for an itinerant \lq single-band' 
ferromagnet, and emphasized its peculiar magnetization dynamics. 
 Their analysis is, however, still phenomenological 
as to the spin-relaxation term which was introduced by hand.

 In this Letter, we present a microscopic calculation of spin torques, 
especially the $\beta$-term and the Gilbert damping, 
in a localized ($s$-$d$) as well as an itinerant (Stoner) models 
for ferromagnetism. 
 We introduce spin-relaxation processes into the (conducting) electron system 
by magnetic (and spin-orbit) impurities 
causing spin-flip and spin-dependent scatterings. 
 Our result shows that both transverse and longitudinal relaxation 
processes contribute to $\alpha$ and $\beta$, 
and that $\alpha \ne \beta$ in general.

 We first consider a localized ($s$-$d$) model. 
 It consists of localized $d$-spins, 
${\bm S}$, and conducting $s$-electrons, which are coupled {\it via} 
the $s$-$d$ exchange interaction 
\begin{eqnarray}
 H_{\rm sd} 
&=& - M \int d {\bm r} \, {\bm n} ({\bm r}) \!\cdot\! 
     \hat {\bm \sigma} ({\bm r}). 
\label{eq:H_sd}
\end{eqnarray} 
 Here we put ${\bm S}=S {\bm n}$ with a unit vector ${\bm n}$ pointing 
in the direction of spin \cite{com0}, 
$\hat {\bm \sigma}({\bm r}) 
= c^\dagger ({\bm r}) {\bm \sigma} c({\bm r})$ 
represents (twice) the $s$-electron spin density, with 
$c^\dagger = (c^\dagger_\uparrow , c^\dagger_\downarrow ) $  
being electron creation operators, 
${\bm \sigma}$ the Pauli spin-matrix vector, 
and $M = J_{\rm sd}S$ with $J_{\rm sd}$ being the $s$-$d$ exchange coupling 
constant. 
 The total Hamiltonian of the system is given by 
$H_{\rm tot} = H_S + H_{\rm el} + H_{\rm sd}$, where 
$H_S$ and $H_{\rm el}$ are for localized $d$-spins and $s$-electrons, 
respectively.

 The dynamics of magnetization, $-{\bm n}$ \cite{com0}, will be 
described by the Landau-Lifshitz-Gilbert (LLG) equation 
\begin{eqnarray}
  \dot {\bm n} &=& \gamma_0 {\bm H}_{\rm eff} \times {\bm n} 
                 + \alpha_0 \dot {\bm n} \times {\bm n} + {\bm t}_{\rm el}' , 
\label{eq:LLG}
\end{eqnarray}
where $\gamma_0{\bm H}_{\rm eff}$ and $\alpha_0$ are an effective field and a 
Gilbert damping constant, respectively, both coming from $H_S$ and not 
from processes involving $s$-electrons. 
 Effects of conducting $s$-electrons are contained in the spin torque 
\begin{eqnarray}
 {\bm t}_{\rm el} ({\bm r}) 
&\equiv&  \frac{\hbar S}{a^3} \, {\bm t}_{\rm el}' ({\bm r})
\ = \  M {\bm n} ({\bm r}) \times 
    \langle \hat {\bm \sigma} ({\bm r}) \rangle_{\rm n.e.}, 
\label{eq:torque0}
\end{eqnarray}
which comes from $H_{\rm sd}$. 
(Here $a^3$ is the volume per localized spin.) 
 The calculation of spin torque is thus equivalent to that of 
$s$-electron spin polarization, 
$\ \langle \hat {\bm \sigma} ({\bm r}) \rangle_{\rm n.e.}$, 
or precisely, its perpendicular projection 
$\langle \hat{\bm \sigma}_\perp ({\bm r}) \rangle_{\rm n.e.}$ to ${\bm n}$ 
\cite{com1}, 
in such nonequilibrium states with current flow and spatially varying 
magnetization (for the $\beta$-term), or with time-dependent magnetization 
(for Gilbert damping). 
 Here and hereafter, $\langle \cdots \rangle_{\rm n.e.}$ represents 
statistical average in such nonequilibrium states.

 Generally, the spin torque is expressed as
\begin{eqnarray}
 {\bm t}_{\rm el} 
&=& a_0 \dot {\bm n} + ({\bm a} \!\cdot\! {\bm \nabla})\,  {\bm n} 
  + b_0 \, ({\bm n} \times \dot {\bm n}) 
  + {\bm n} \times ({\bm b} \cdot\! {\bm \nabla}) \, {\bm n}  \ \ 
\label{eq:torque1} 
\end{eqnarray}
in the first order in time derivative and spatial gradients. 
 The ${\bm a}$-term includes the spin-transfer torque, 
$- ({\bm v}_{\rm s} \!\cdot\! {\bm \nabla}){\bm n}$,
where 
${\bm v}_{\rm s} = - (a^3/2eS) {\bm j}_{\rm s}$ 
($e>0$ being the elementary charge) \cite{com0} 
is a quantity which may be called as the spin-transfer velocity. 
 (The $a_0$-term just renormalizes the magnitude of spin, 
 $S \to S - (a^3/\hbar) \, a_0$ on the left-hand side of the LLG equation, 
 or in front of the kinetic (\lq\lq Berry phase'') term in the spin 
 Lagrangian \cite{STK05}.) 
 The ${\bm b}$-term is the central issue in the present study. 
 We also focus on $b_0$, which gives damping of the Gilbert type, 
with dimensionless damping constant $\alpha = - (a^3/\hbar S) b_0$. 
 The ${\bm b}$-term is expressed as 
$- \beta \, {\bm n} \times ({\bm v}_{\rm s} \!\cdot\! {\bm \nabla}) \, {\bm n}$
(in ${\bm t}_{\rm el}'$) \cite{com0} 
in the literature \cite{Thiaville05,Barnes05,TBB}, 
which defines the constant $\beta$.

 The spin torque (\ref{eq:torque1}) corresponds to the $s$-electron 
spin polarization $\langle \hat{\bm \sigma}_\perp \rangle_{\rm n.e.}$ 
\cite{com1} given by 
\begin{eqnarray}
  \frac{1}{M} \left[\, 
    b_0 \dot {\bm n} + ({\bm b} \!\cdot\! {\bm \nabla})\,  {\bm n} 
  - a_0 \, ({\bm n} \times \dot {\bm n}) 
  - {\bm n} \times ({\bm a} \cdot\! {\bm \nabla}) \, {\bm n} \,\right] . 
\label{eq:sigma_perp} 
\end{eqnarray}
 To calculate the coefficients, $a_\mu$ and $b_\mu$, 
we follow TBB \cite{TBB} (see also Ref.\cite{TFH}) and consider a small 
transverse fluctuation, 
${\bm u} = (u^x,u^y,0)$, $|{\bm u}| \ll 1$, 
around a uniformly magnetized state, ${\bm n} = \hat z$, such that 
${\bm n} = \hat z + {\bm u}$. 
 In the \lq unperturbed' state, ${\bm n} = \hat z$, the $s$-electrons 
are described by the Hamiltonian 
\begin{eqnarray}
  {\cal H}_0 
&=& \sum_{{\bm k} \sigma} 
   (\varepsilon_{{\bm k}} - \varepsilon_{{\rm F} \sigma}) \, 
   c^\dagger_{{\bm k} \sigma} c^{\phantom{\dagger}}_{{\bm k} \sigma} 
   + V_{\rm imp} , 
\label{eq:H0}
\end{eqnarray}
and have a spin polarization 
$ \langle \hat {\bm \sigma} \rangle_0 = \rho_{\rm s} \hat z $. 
 Here $\rho_{\rm s}  = n_\uparrow - n_\downarrow$, 
$n_\sigma = k_{{\rm F} \sigma}^3/6\pi^2$, 
$\varepsilon_{{\rm F} \sigma} 
 = \hbar^2 k_{{\rm F} \sigma}^2/2m 
 = \varepsilon_{\rm F} + \sigma M$, and $V_{\rm imp}$ is 
the impurity potential specified later (Eq.(\ref{eq:Vimp})).  
 The subscript $\sigma = \uparrow, \downarrow$ 
corresponds, respectively, to $\sigma = +1, -1$ 
in the formula 
(and to $\bar\sigma = \downarrow, \uparrow$ or $-1, +1$) . 
 In the presence of 
${\bm u}({\bm r},t) 
= {\bm u}({\bm q},\omega) \, {\rm e}^{i({\bm q}\cdot{\bm r}-\omega t)}$, 
the $s$-electrons feel a perturbation 
(note that $H_{\rm el} + H_{\rm sd} = {\cal H}_0 + {\cal H}_1$) 
\begin{eqnarray}
  {\cal H}_1  
&=& - M \sum_{{\bm k} \sigma} c^\dagger_{{\bm k}+{\bm q}} {\bm \sigma} 
                     c^{\phantom{\dagger}}_{{\bm k}} 
           \!\cdot\! {\bm u} ({\bm q},\omega) \, {\rm e}^{-i\omega t} , 
\label{eq:H1}
\end{eqnarray}
and acquires a transverse component \cite{com1} 
\begin{eqnarray}
  \langle \hat \sigma_\perp^{\prime \alpha} (x) \rangle_{\rm n.e.} 
&=& M \int_{-\infty}^t dt' \int d{\bm r}' 
   \chi_\perp^{\alpha \beta} (x-x') \, u^\beta (x') 
\end{eqnarray}
in their spin polarization, in the first order in ${\bm u}$. 
 Here, $x=({\bm r},t)$, $x'=({\bm r}',t')$, and $\chi_\perp^{\alpha \beta}$ 
is the transverse spin susceptibility in a uniformly magnetized state. 
($\alpha, \beta = x,y$ specify the transverse componentsC
and summing over $\beta$ is implied.) 
 Writing in Fourier components, 
\begin{eqnarray}
  \langle \hat \sigma_\perp^{\prime \alpha} ({\bm q}, \omega ) \rangle_{\rm n.e.} 
&=& M \chi_\perp^{\alpha \beta} ({\bm q}, \omega + i0) \, 
    u^\beta ({\bm q},\omega) , 
\end{eqnarray}
we expand $\chi_\perp^{\alpha \beta} ({\bm q}, \omega + i0)$ as 
\begin{eqnarray}
&{}& 
    \chi_\perp^{\alpha \beta} ({\bm q}, \omega + i0) 
  - \chi_\perp^{\alpha \beta} ({\bm 0}, 0) 
\nonumber \\ 
&=& \frac{1}{M^2} \left[\, 
     i({\bm b} \!\cdot\! {\bm q} - b_0 \omega ) \delta_{\alpha \beta} 
   + i({\bm a} \!\cdot\! {\bm q} - a_0 \omega ) \varepsilon_{\alpha \beta} 
     \,\right]
\label{eq:chi_ab}
\end{eqnarray}
up to the first order in ${\bm q}$ and $\omega$. 
(Here $\varepsilon_{\alpha\beta}$ is an antisymmetric tensor in 2D, 
with $\varepsilon_{xy} = 1$.) 
 Below we will see that 
$M \chi_\perp^{\alpha\beta}({\bm 0}, 0) = \rho_{\rm s} \delta_{\alpha\beta}$. 
 Therefore, 
$ \langle \hat {\bm \sigma} \rangle_{\rm n.e.} 
= \rho_{\rm s} {\bm n} + \langle \hat {\bm \sigma}_\perp \rangle_{\rm n.e.}$, 
with $\langle \hat {\bm \sigma}_\perp \rangle_{\rm n.e.}$ 
given by 
\begin{eqnarray}
  \frac{1}{M} \left[\, 
b_0 \dot {\bm u} + ({\bm b} \!\cdot\! {\bm \nabla}) {\bm u} 
   - a_0 (\hat z \times \dot {\bm u}) 
   - \hat z \times ({\bm a} \!\cdot\! {\bm \nabla}) {\bm u}  \,\right] , 
\label{eq:sigma_u}
\end{eqnarray}
which coincides with (\ref{eq:sigma_perp}) within the present accuracy 
({\it i.e.}, neglecting 
${\bm u} \times \dot {\bm u} = {\cal O}(u^2)$), 
and leads to the spin torque (\ref{eq:torque1}). 
 The problem thus reduces to the calculation of the coefficients, 
$a_\mu$ and $b_\mu$, in Eq.(\ref{eq:chi_ab}) \cite{Sakuma}.

 We consider a 3D electron system with 
$\varepsilon_{\bm k} = \hbar^2 {\bm k}^2/2m$, 
which is affected by the impurity potential represented by 
\begin{eqnarray}
 V_{\rm imp} 
&=&  u \sum_i \delta ({\bm r} - {\bm R}_i) 
   + u_{\rm s} \sum_j {\bm S_j} \!\cdot\! {\bm \sigma} 
               \delta ({\bm r} - {\bm R}_j') 
\label{eq:Vimp}
\end{eqnarray}
in the first-quantization form. 
 The first term describes potential scattering, and the second describes 
spin scattering by impurity spins, ${\bm S_j}$. 
 We take a quenched average for the impurity spin direction as 
\begin{eqnarray}
  \overline{S_i^\alpha S_j^\beta} 
&=& \delta_{ij} \delta_{\alpha\beta} \times \left\{ \begin{array}{cc} 
    \overline{S_\perp^2} & (\alpha, \beta = x,y) \\ 
    \overline{S_z^2}     & (\alpha, \beta = z) 
    \end{array} \right.
\end{eqnarray}
as well as for the impurity positions, ${\bm R}_i$ and ${\bm R}_j'$, 
as usual. 
 The electron damping rate is then given by 
\begin{eqnarray}
  \gamma_\sigma  
 =  \frac{1}{2\tau_\sigma} 
 =  \pi n_{\rm i}u^2  \nu_\sigma  
  + \pi n_{\rm s} u_{\rm s}^2 
    ( 2\overline{S_\perp^2} \nu_{\bar\sigma} + \overline{S_z^2} \nu_\sigma ) 
\label{eq:gamma}
\end{eqnarray}
in the first Born approximation, 
where $n_{\rm i}$ ($n_{\rm s}$) is the concentration of normal (magnetic) 
impurities, and 
$\nu_\sigma = m \, k_{{\rm F} \sigma}/2\pi^2\hbar^2$ is the 
density of states (DOS) at $\varepsilon_{{\rm F} \sigma}$. 
 We can include the spin-orbit impurities, of scattering amplitude 
$ iu_{\rm s.o.} ({\bm k} \times {\bm k}') \!\cdot\! {\bm \sigma}$, 
by adding 
$n_{\rm s.o.} u_{\rm s.o.}^2 \overline{({\bm k} \times {\bm k}')_i^2}$ 
to 
$n_{\rm s} u_{\rm s}^2 \overline{S_i^2}$, in all expressions 
in the present study.

 We assume that $\gamma_\sigma \ll \varepsilon_{{\rm F} \sigma}$ 
and $\gamma_\sigma \ll M$, and calculate $\alpha$ and $\beta$ 
in the lowest non-trivial order in 
$x_\sigma \equiv \gamma_\sigma /4\varepsilon_{{\rm F} \sigma}$ and 
$\gamma_\sigma / M$, both being collectively denoted as $\gamma$. 
 We will see that $\alpha$ and $\beta$ are quantities of 
${\cal O}(\gamma )$. 
 For a perturbative estimate in $\gamma$, only the first-order correction 
is sufficient to the $\sigma^\alpha$- (and $\sigma^\beta$-) vertex, 
since the corresponding \lq diffusion' ladder has a large mass $\sim M$.

\begin{figure}[b]
  \begin{center}
  \includegraphics[scale=0.4]{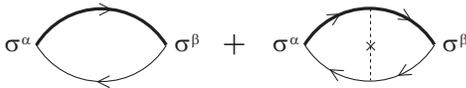}
  \vskip -2mm
  \end{center}
\caption{Transverse spin susceptibility up to ${\cal O}(\gamma )$. 
The dotted line with a cross represents scattering 
by impurities, either non-magnetic or magnetic 
(including spin-orbit). 
The thick (thin) solid line carries Matsubara frequency 
$i\varepsilon_n + i\omega_\lambda$ ($i\varepsilon_n$).}
\end{figure}

{\bf Gilbert damping :}
 We first study the $\omega$-linear terms in the uniform (${\bm q}={\bm 0}$) 
part of the transverse spin susceptibility, 
$\chi_\perp^{\alpha \beta} ({\bm q}={\bm 0}, \omega + i0)$. 
 Contributions up to ${\cal O}(\gamma)$, represented in Fig.1, 
are given, in Matsubara form, by 
\begin{eqnarray}
  \chi_\perp^{\alpha \beta} ({\bm 0}, i\omega_\lambda ) 
&=& - T \sum_n \sum_{\sigma} 
    (\delta_{\alpha \beta} + i\sigma \varepsilon_{\alpha \beta}) \, 
    \varphi_\sigma , 
\label{eq:chi_ab_phi}
\end{eqnarray}
where
\begin{eqnarray}
  \varphi_\sigma
&\equiv& 
  \varphi_\sigma (i\varepsilon_n ; i\omega_\lambda ) 
\ = \  \bar\chi_\sigma 
   + \tilde\Gamma_0 \bar\chi_\sigma^2 , 
\end{eqnarray}
with
$ \bar\chi_\sigma
 \equiv \bar\chi_\sigma (i\varepsilon_n ; i\omega_\lambda ) 
 = \sum_{{\bm k}} G_{{\bm k} \bar\sigma}(i\varepsilon_n + i\omega_\lambda)
                  G_{{\bm k} \sigma}(i\varepsilon_n) $, 
$G_{{\bm k} \sigma}(z) 
= (z - \varepsilon_{\bm k} + \varepsilon_{{\rm F} \sigma} + i\gamma_\sigma 
 {\rm sgn} ({\rm Im} z))^{-1}$. 
 The vertex correction is associated with a factor 
$ \tilde\Gamma_0
 =  n_{\rm i} u^2 - n_{\rm s} u_{\rm s}^2 \overline{S_z^2}$, 
where the $S_z$ scattering changes sign 
whereas $S_x$ and $S_y$ scatterings cancel each other. 
 After analytic continuation ($i\omega_\lambda \to \omega + i0$) 
and extracting the $\omega$-linear terms, we have 
\begin{eqnarray}
 b_0 
&=& \frac{M^2\hbar}{2\pi} \sum_\sigma 
    {\rm Re} \{ \varphi^{(1)}_\sigma (0;0) - \varphi^{(2)}_\sigma (0;0) \}, 
\label{eq:b_0_phi}
\end{eqnarray}
at absolute zero, $T=0$. 
 Here the upper labels, (1) and (2), on $\varphi_\sigma$ 
(and $\bar\chi_\sigma$ below) indicate the analytic continuations, 
$G(i\varepsilon_n +i\omega_\lambda) G(i\varepsilon_n) \to G^{\rm R}G^{\rm R}$ 
and $G^{\rm R} G^{\rm A}$, respectively \cite{com3}.

 By an explicit evaluation of the ${\bm k}$-integrals, we have 
$ \bar\chi^{(1)}_\sigma (0;0) 
 =  (i\pi/2M)[ \nu_- +iy -i (\nu_- \gamma_-/M)] + {\cal O}(\gamma^2)$, 
$ \bar\chi^{(2)}_\sigma (0;0) 
 = - (i\pi/2M)[ \sigma\nu_+ -iy +i (\nu_+ \gamma_+/M)] + {\cal O}(\gamma^2)$, 
and thus 
$ b_0 = - \hbar ( \nu_+ \gamma_+ - \nu_- \gamma_- )/4 
        + (\pi\hbar\tilde\Gamma_0 /8) \, (\nu_+^2 - \nu_-^2 ) $. 
 Here $\nu_\pm = \nu_\uparrow \pm \nu_\downarrow$, 
$\gamma_\pm = (\gamma_\uparrow \pm \gamma_\downarrow )/2$, 
and $y = \nu_\uparrow x_\uparrow - \nu_\downarrow x_\downarrow$. 
 Using Eq.(\ref{eq:gamma}), we finally obtain 
\begin{eqnarray}
  \alpha 
&=&  \pi n_{\rm s} u_{\rm s}^2 \left[\, 
     2 \overline{S_z^2} \nu_\uparrow \nu_\downarrow 
    + \overline{S_\perp^2} (\nu_\uparrow ^2 + \nu_\downarrow^2 ) 
    \right] 
    \times \frac{a^3}{S} . 
\label{eq:alpha1}
\end{eqnarray}
 As expected, only the spin (and spin-orbit) scattering contributes to 
$\alpha$,  
and the potential scattering ($\sim n_{\rm i}u^2$) does not 
thanks to the cancellation between selfenergy and vertex correction 
\cite{com4}.

 By a similar analysis, we obtain $a_0 = - \hbar \rho_{\rm s}/2$. 

 For $i\omega_\lambda = 0$, $\varphi_\sigma$ is independent of $\sigma$. 
 Thus, from Eq.(\ref{eq:chi_ab_phi}), $\chi_\perp^{\alpha \beta}({\bm 0},0)$ 
is proportional to $\delta_{\alpha\beta}$. 
 Explicit calculation shows that 
$ \chi_\perp^{\alpha \beta}({\bm 0}, 0) = (\rho_{\rm s}/M) 
\, \delta_{\alpha\beta}$ 
as used above.

{\bf $\beta$-term :}
 We next examine the ${\bm q}$-linear terms in the presence of current flow 
under static ${\bm n}({\bm r}) = \hat z + {\bm u}({\bm r})$. 
 We produce a current-carrying state by applying a d.c. electric field
${\bm E}$, and calculate a linear response of 
$\sigma_\perp^\alpha$ to ${\bm E}$. 
 In a similar way to the Kubo formula for electrical conductivity 
\cite{Kubo}, one can derive 
\begin{eqnarray}
&{}& 
  \langle \hat\sigma^\alpha_\perp ({\bm q}) \rangle_{\rm n.e.} 
 \ = \  \chi_i^\alpha ({\bm q}) E_i , 
\\
&{}& \ \ 
 \chi_i^\alpha ({\bm q}) 
\ = \ \lim_{\omega \to 0}
    \frac{ K_i^\alpha ({\bm q},\omega +i0) -K_i^\alpha ({\bm q},0)}{i\omega},
\label{eq:linear72}
\\
&{}& \ \ 
 K_i^\alpha ({\bm q}, i\omega_\lambda ) 
\ = \ \int_0^\beta d\tau \, {\rm e}^{i\omega_\lambda \tau} \, 
    \langle \, {\rm T}_\tau \, \sigma^\alpha ({\bm q}, \tau) 
            \, J_i \, \rangle . \ \ \ 
\label{eq:K72}
\end{eqnarray}
 Here ${\bm J} = - e \sum_{{\bm k}} {\bm v}_{\bm k} 
   \, c_{{\bm k} \sigma}^\dagger c_{{\bm k} \sigma}^{\phantom{\dagger}}$ 
is the total current (${\bm v}_{\bm k} = \hbar {\bm k}/m$), and  
the average $\langle \cdots \rangle$ is taken in the thermal equilibrium 
state determined by ${\cal H}_0 + {\cal H}_1$ 
with $\omega=0$ (in Eq.(\ref{eq:H1})).  
 To have a non-vanishing contribution, we need to extract 
$u^\beta$ and $q_j$ both in first order, and put 
\begin{eqnarray}
  K_i^{\alpha} ({\bm q}, i\omega_\lambda ) 
&=& - eM K_{ij}^{\alpha\beta} (i\omega_\lambda ) q_j u^\beta({\bm q},0). 
\end{eqnarray}
 Diagrammatic expressions for $K_{ij}^{\alpha\beta}$ are given in Fig.2.

\begin{figure}[b]
  \begin{center}
  \includegraphics[scale=0.32]{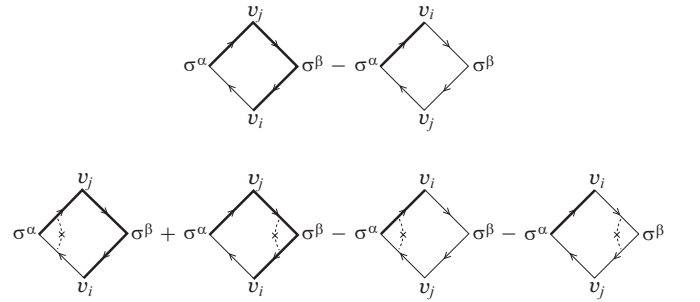}
  \vskip 0mm
  \end{center}
\caption{ 
 The $E_i$-, $q_j$- and $u^\beta$-linear coefficient, 
$K_{ij}^{\alpha\beta}$, of the transverse spin polarization, 
$\langle \hat\sigma^\alpha_\perp ({\bm q}) \rangle_{\rm n.e.}$, 
in the presence of current flow ($E_i$) and magnetization texture 
($q_j u^\beta$). 
 The velocity vertices with $v_i$ and $v_j$ are associated with 
$E_i$ and $q_j$, respectively. 
 Other graphical meanings are the same as Fig.1. }
\label{fig:susceptibility_current_uni}
\end{figure}

 The calculation of ${\bm b}$ (and ${\bm a}$) as a linear response to 
${\bm E}$ is similar 
to the calculation of transport coefficients, but requires a closer analysis 
to collect all the terms one order higher in $\gamma$. 
 The result, up to ${\cal O}(\gamma^0)$, is 
\begin{eqnarray}
  {\bm a} 
&=& \frac{e\hbar {\bm E}}{2m} 
     \sum_\sigma \sigma n_\sigma \tau_\sigma 
\ = \  \frac{\hbar}{2e} \sigma_{\rm s} {\bm E} 
\ = \  \frac{\hbar}{2e} {\bm j}_{\rm s} , 
\label{eq:a}
\\
  {\bm b} 
&=&  \frac{e\hbar {\bm E}}{2mM} \left[ 
      \sum_\sigma \sigma n_\sigma  \tau_\sigma \gamma_{\bar\sigma} 
    - \pi \tilde\Gamma_0 \, 
      \sum_\sigma \sigma n_\sigma \tau_\sigma \nu_{\bar\sigma} 
      \right] 
\label{eq:b'}
\\ 
  &=&  \frac{\pi n_{\rm s}u_{\rm s}^2}{M} \!\cdot\!  
     \frac{\hbar}{2e} \left[ 
    \bigl( \overline{S_\perp^2} + \overline{S_z^2} \bigr) \nu_+ 
    {\bm j}_{\rm s} 
  + \bigl( \overline{S_\perp^2} - \overline{S_z^2} \bigr) \nu_- 
    {\bm j}_{\rm c}
    \right] , \ \ \ \ \ \ 
\label{eq:b}
\end{eqnarray}
where 
${\bm j}_{\rm s} = \sigma_{\rm s}{\bm E}
 = {\bm j}_\uparrow - {\bm j}_\downarrow $ is the spin current, and 
${\bm j}_{\rm c} = \sigma_{\rm c}{\bm E} 
 = {\bm j}_\uparrow + {\bm j}_\downarrow $ is the charge current, 
with $\{ \sigma_{\rm c}, \sigma_{\rm s} \} = (e^2/m) 
(n_\uparrow \tau_\uparrow \pm n_\downarrow \tau_\downarrow)$ 
being charge and spin conductivities. 
 Similarly to the case of $\alpha$, only the spin (and spin-orbit) scattering 
contributes to ${\bm b}$ \cite{com4}. 
 We also see that ${\bm b}$ is mainly determined by ${\bm j}_{\rm s}$, but 
also depends on ${\bm j}_{\rm c}$ for a general case of 
$\overline{S_\perp^2} \ne \overline{S_z^2}$, 
and that 
\begin{eqnarray}
  \beta 
 &=&  \frac{\pi n_{\rm s}u_{\rm s}^2}{M} 
    \left[ 
    \bigl( \overline{S_\perp^2} + \overline{S_z^2} \bigr) \nu_+ 
  + \frac{1}{P_j} 
    \bigl( \overline{S_\perp^2} - \overline{S_z^2} \bigr) \nu_- 
    \right] , 
\label{eq:beta1}
\end{eqnarray}
where $P_j = \sigma_{\rm s} / \sigma_{\rm c} $ 
is the polarization of the current.

 In terms of longitudinal ($\tau_L^{\phantom{1}}$) and transverse 
($\tau_T^{\phantom{1}}$) spin-relaxation times, obtained as 
$\tau_L^{-1} = 4 \pi n_{\rm s} u_{\rm s}^2 \overline{S_\perp^2} \nu_+ / \hbar$ 
and 
$\tau_T^{-1} = 2 \pi n_{\rm s} u_{\rm s}^2 
    \bigl( \overline{S_\perp^2} + \overline{S_z^2} \bigr) \nu_+ / \hbar$ 
\cite{com5}, 
we have 
\begin{eqnarray}
  \alpha 
 &=&  \frac{a^3 \hbar \nu_+}{4S} \left[ 
    \left( 1 - P_\nu^2 \right) \frac{1}{\tau_T^{\phantom{1}}} 
    + P_\nu^2 \frac{1}{\tau_L^{\phantom{1}}}   \right] , 
\label{eq:alpha2}
\\
  \beta 
 &=&  \frac{\hbar}{2M} \left[ 
    \left( 1 - \frac{P_\nu}{P_j} \right) \frac{1}{\tau_T^{\phantom{1}}} 
    + \frac{P_\nu}{P_j} \frac{1}{\tau_L^{\phantom{1}}}   \right] , 
\label{eq:beta2}
\end{eqnarray}
where $P_\nu = \nu_- / \nu_+ $ is the DOS asymmetry. 
 We see that 
both longitudinal and transverse spin-relaxation processes contribute 
to $\alpha$ and $\beta$ in general ({\it i.e.}, if $P_\nu \ne 0$), 
in contrast to the demonstration of Ref.\cite{TBB}. 
 For \lq\lq isotropic'' impurities with 
$\overline{S_\perp^2} = \overline{S_z^2}$ 
and thus 
$\tau_L^{\phantom{1}} = \tau_T^{\phantom{1}} \equiv \tau_{\rm s}$, 
we have 
\begin{eqnarray}
  \alpha 
 &=&  \frac{a^3 \nu_+}{4S} \!\cdot\! \frac{\hbar}{\tau_{\rm s}} , \ \ \ \ \  
  \beta 
\ = \  \frac{\hbar}{2M\tau_{\rm s}} . 
\end{eqnarray}
 The present results are obtained from those of Zhang and Li \cite{Zhang05} 
if we identify their $\tau_{sf}$ and $n_0$ with our  $\tau_{\rm s}$ 
(or $\tau_T^{\phantom{1}}$ when $P_\nu = 0$) and $M\nu_+$, respectively.

Finally, we consider an itinerant single-band model such as the Stoner model 
treated in a mean-field approximation. 
 Suppose we apply a small transverse field, 
${\bm h} = {\bm h}_{{\bm q}, \omega} 
 {\rm e}^{i({\bm q} \cdot {\bm r} - \omega t)}$, 
on a uniform ferromagnetic state. 
 The corresponding magnetization dynamics is described by 
$ u^\alpha ({\bm q}, \omega ) 
= - [ \, \chi_\perp^{{\rm RPA}} ({\bm q}, \omega ) \, ]^{\alpha\beta} 
    h_{{\bm q}, \omega}^\beta$ , 
within linear response. 
 Here 
${\bm u} \equiv \langle \hat {\bm \sigma}_\perp' \rangle_{\rm n.e.}$ now, 
and $\chi_\perp^{\rm RPA} \equiv (\chi_\perp^{-1} - J)^{-1}$ 
is the RPA (transverse) susceptibility, with $\chi_\perp$ given by 
Eq.(\ref{eq:chi_ab}) and $J$ being the ferromagnetic exchange coupling 
constant. 
 We write 
$ (\chi_\perp^{\rm RPA})^{-1} {\bm u} = - {\bm h}$, 
and expand as 
$ \chi_\perp^{-1} 
= \chi_{\perp,0}^{-1} 
- \chi_{\perp,0}^{-1} \chi_{\perp,1}^{\phantom{1}} \chi_{\perp,0}^{-1} $ 
up to ${\cal O} (q, \omega)$, 
where $\chi_{\perp,n}$ represents terms $n$-th order 
in ${\bm q}$ and $\omega$. 
 Using the self-consistent equation, $M = J \rho_{\rm s}$, at equilibrium, 
we obtain 
\begin{eqnarray} 
 \rho_{\rm s}^{-2} [ \, 
   b_0 \dot {\bm u} 
 + ({\bm b} \cdot\! {\bm \nabla}) {\bm u}
 - \hat z \times ( 
   a_0 \dot {\bm u} + ({\bm a} \!\cdot\! {\bm \nabla})\, {\bm u} ) 
 \, ]
&=&  {\bm h} .  \ \ \ 
\label{eq:LLG_Stoner1} 
\end{eqnarray}
 Within ${\cal O}(u)$, this is consistent with the LLG equation 
\begin{eqnarray} 
 \dot {\bm n} 
&=& ({\bm a}' \!\cdot\! {\bm \nabla})\, {\bm n} 
  + {\bm n} \times [ \,  b_0'  \dot {\bm n} 
  + ({\bm b}' \cdot\! {\bm \nabla}) {\bm n} 
  -  \rho_{\rm s}^2 {\bm h}' \, ] . 
\label{eq:LLG_Stoner2} 
\end{eqnarray}
 Primed quantities here are those divided by 
$- a_0 = \hbar \rho_{\rm s} / 2$. 
 This amounts to replacing $2S$  in the localized model 
by $\rho_{\rm s} a^3$.  
 For \lq\lq isotropic'' impurities, we have  
$\alpha = \hbar \nu_+ / 2\rho_{\rm s} \tau_{\rm s}$ 
and $\beta = \hbar / 2M \tau_{\rm s}$, and thus 
$\alpha \ne \beta$ even in an itinerant single-band model, 
in contrast to the result of Ref.\cite{TBB}.

 We have presented a model calculation of spin torques, 
especially the Gilbert damping and the so-called $\beta$-term, 
on the basis of two types of microscopic models for ferromagnetism 
(localized and itinerant) 
and a controlled approximation. 
 Magnetic (and spin-orbit) impurities have been used as a model for spin 
relaxation in conducting electron systems. 
 In actual systems, however, the origin of spin relaxation 
will be various depending on specific systems. 
 We hope the present formalism will be useful for subsequent studies 
treating such more realistic models.

We would like to thank Y. Nakatani, H. Ohno, T. Ono and  Y. Suzuki 
for valuable discussions. 
 This work is financially supported by Monka-sho, Japan.


\begin{thebibliography}{99}
\bibitem{com0}
 Please note that the direction of ${\bm n}$ is opposite to the magnetization, 
and the direction of electron flow is opposite to that of spin and charge 
currents, ${\bm j}_{\rm s}$ and ${\bm j}_{\rm c}$, which are defined 
just after Eq.(\ref{eq:b}).
\bibitem{com1}
 We define 
$\hat{\bm \sigma}_\perp 
= \hat{\bm \sigma}-{\bm n}({\bm n} \!\cdot\! \hat{\bm \sigma})$ 
and 
$\hat{\bm \sigma}_\perp' 
= \hat{\bm \sigma}-\hat z (\hat z \!\cdot\! \hat{\bm \sigma})$. 
 Note that $\langle \hat {\bm \sigma} \rangle_{\rm n.e.}$ in 
Eq.(\ref{eq:torque0}) can be replaced by 
$\langle \hat {\bm \sigma}_\perp \rangle_{\rm n.e.}$.
\bibitem{com3}
 In deriving Eq.(\ref{eq:b_0_phi}), we have noted that, 
under $\sigma \to -\sigma$, 
$\varphi^{(1)}_\sigma (0;0)$
and ${\rm Re} \, \varphi^{(2)}_\sigma (0;0)$ 
are even, whereas  
${\rm Im} \varphi^{(2)}_\sigma (0;0)$ and 
$ \partial \varphi^{(1)}_\sigma (\varepsilon ; \omega ) 
 / \partial \omega  |_{\omega = 0} $ 
are odd.
\bibitem{com4}
 Note that spin-orbit coupling in the bulk band structure is not 
taken into account in the present study.
\end{thebibliography}
\end{document}